\documentstyle[12pt]{article}
\newcommand{\bstar}{\begin{petit} \noindent {\Large $\star$} }
\newcommand{\estar}{\end{petit}}
\newcommand{\journal}[4]{{\em #1~}#2\,(19#3)\,#4;}

\newcommand{\hpa}{\journal {Helv. Phys. Acta}}

\newcommand{\ijmp}{\journal {Int. J. Mod. Phys.}}

\newcommand{\pr}{\journal {Phys. Rev.}}

\newcommand{\prl}{\journal {Phys. Rev. Lett.}}

\newcommand{\cmp}{\journal {Commun. Math. Phys.}}

\newcommand{\np}{\journal {Nucl. Phys.}}
\newcommand{\pl}{\journal {Phys. Lett.}}

\newcommand{\nc}{\journal {Nuovo Cimento}}

\renewcommand{\a}{\alpha}

           \newcommand{\G}{\Gamma}
\renewcommand{\d}{\delta}         \newcommand{\D}{\Delta}

           \renewcommand{\S}{\Sigma}
         
\newcommand{\f}{{\phi}}           
\newcommand{\vf}{{\varphi}}

\newcommand{\FF}{{\cal F}}
\newcommand{\GG}{{\cal G}}

\newcommand{\SS}{{\cal S}}

\newcommand{\Sla}{\raise.15ex\hbox{$/$}\kern -.70em D}

\newcommand{\lp}{\left(}\newcommand{\rp}{\right)}
\newcommand{\lc}{\left[}\newcommand{\rc}{\right]}
\newcommand{\lac}{\left\{}\newcommand{\rac}{\right\}}

\newcommand{\tr}{{\rm {Tr} \,}}
\newcommand{\half}{\frac 1 2}

\newcommand{\dfud}[2]{{\displaystyle{\frac{\delta #1}{\delta #2}}}}
\newcommand{\dfrac}[2]{{\displaystyle{\frac{#1}{#2}}}}

\newcommand{\dint}{\displaystyle{\int}}

\newcommand{\sla}{\raise.15ex\hbox{$/$}\kern -.57em}
\newcommand{\twiddle}{\lower.9ex\rlap{$\kern -.1em\scriptstyle\sim$}}

\newcommand{\es}{\\[3mm]}
\newcommand{\equ}[1]{(\ref{#1})}
\newcommand{\eq}{\begin{equation}}
\newcommand{\eqn}[1]{\label{#1}\end{equation}}
\newcommand{\eea}{\end{eqnarray}}
\newcommand{\eqa}{\begin{eqnarray}}
\newcommand{\eqan}[1]{\label{#1}\end{eqnarray}}
\newcommand{\ba}{\begin{array}}
\newcommand{\ea}{\end{array}}
\newcommand{\eqac}{\begin{equation}\begin{array}{rcl}}
\newcommand{\eqacn}[1]{\end{array}\label{#1}\end{equation}}

\makeatletter
\@addtoreset{equation}{section}
\makeatother

\newcommand{\Bb}{{\bar B}}
\newcommand{\cb}{{\bar c}}
\newcommand{\Ab}{{\bar A}}
\newcommand{\Db}{{\bar D}}
\newcommand{\dv}{{\dint dV\,}}
\newcommand{\ds}{{\dint dS\,}}
\newcommand{\dsb}{{\dint d\bar S\,}}
\newcommand{\ssg}{{\SS_\G}}
\newcommand{\ap}{{\dot\a}}

\begin{document}
\noindent


\begin{center}

{\huge The Antighost Equation in $N=1$ Super-Yang-Mills Theories}
\vspace{8mm}

{\Large Olivier Piguet}\footnote{On leave of
absence from:
{\it D\'epartement de Physique Th\'eorique --
     Universit\'e de Gen\`eve, 24 quai E. Ansermet -- CH-1211 Gen\`eve
     4, Switzerland.}}$^,$\footnote{Supported in part by the
     Swiss National Science Foundation and the Brazilian National Research
     Council (CNPq)}
\vspace{3mm}

{\it Instituto da F{\'\i}sica, Universidade Cat\'olica de Petr\'opolis\\
25610-130 Petr\'opolis, RJ, Brazil\\
\vspace{.5mm}

and\\
\vspace{.5mm}

Centro Brasileiro de Pesquisas F{\'\i}sicas (CBPF)\\
Rua Xavier Sigaud 150, 22290-180 Urca, RJ, Brazil}
\vspace{10mm}

{\Large and Silvio P. Sorella}
\vspace{3mm}

{\it UERJ \\
 Departamento de F{\'\i}sica Te{\'o}rica \\
 Instituto de F{\'\i}sica, Uerj \\
 Rua S{\~a}o Francisco Xavier, 528 \\
 20550-013, Rio de Janeiro, Brazil \\
 \vspace{.5mm}

and\\
\vspace{.5mm}

Centro Brasileiro de Pesquisas F{\'\i}sicas (CBPF)\\
Rua Xavier Sigaud 150, 22290-180 Urca, RJ, Brazil}
\vspace{3mm}

CBPF--NF--072/95\\
UGVA--DPT 1995/10--904\\
hep-th/9510089
\end{center}
\vspace{7mm}

\begin{center}
{\large {\bf Abstract}}
\end{center}

{\it The antighost equation valid for usual gauge theories in the Landau
gauge, is generalized to the case of $N=1$ supersymmetric gauge theories
in a supersymmetric version of the Landau gauge. This equation, which
expresses the nonrenormalization of the Faddeev-Popov ghost field, plays
an important role in the proof of the nonrenormalization theorems for
the chiral anomalies.}
\newpage

\section{Introduction and Conclusions}
The nonrenormalization theorems for the chiral
anomalies~\cite{ab,bms-nr,lps-nr,p-sor-book}
in ordinary gauge theories heavily rely
on the  (ultraviolet) finiteness of the anomaly operator, more precisely
on the vanishing of its anomalous dimension.
It has been shown that the latter property
is linked to a so-called {\it antighost equation}~\cite{bps,p-sor-book},
valid in the Landau
gauge, expressing
the absence of anomalous dimension
for the ghost field $c$, and more generally, for any invariant polynomial of
$c$ without derivatives. This has allowed for a
simple, general and renormalization scheme independent
proof of the nonrenormalization theorems~\cite{lps-nr,p-sor-book}.
Althoug the antighost equation holds only in the Landau gauge,
the ensuing nonrenormalization theorem is valid in any gauge due to the
gauge independence of the anomaly.

There is an analogous nonrenormalization theorem in supersymmetric
gauge theories (see Appendix A of~\cite{lps-beta-hpa}). As in the usual
theories, one of the ingredients of the proof is the zero anomalous
dimension of the
ghost fields and of their polynomials, which is equivalent to the
finiteness of the diagrams involving  ghost $c$ external lines
and insertions which are polynomial in $c$.
However, in order to prove this finiteness, use has been made of the
``supersymmetric nonrenormalization theorem''
stating the finiteness of the chiral vertices
(see~\cite{chiralvert} and p.126 of ~\cite{ps-book}), which may
be applied here since $c$ is a
chiral superfield.
Since the supersymmetric nonrenormalization theorem assumes the superspace
Feynman rules of exact supersymmetry,
it follows that this proof of the
nonrenormalization theorem for the chiral anomalies
is valid only for theories where the supersymmetry is unbroken. But it is
a fact that all physically interesting theories have explicitely and
softly broken supersymmetry~\cite{susybreak},
and also that, due to infrared effects,
 such a breakdown naturally appears in the
renormalization  of supersymmetric theories with a supersymmetric
gauge fixing~\cite{ps-book}.
Hence it is desirable to extend the proof to these cases.

The aim of the present paper is to derive the
generalization of the antighost equation to the case of supersymmetric
gauge theories quantized in a supersymmetric extension of the Landau
gauge, and to indicate how it leads to a nonrenormalization theorem for
the chiral anomalies which is valid in the broken symmetry case.

\section{BRS Transformations and Classical Action}
The notations and conventions are those of~\cite{ps-book}. The gauge group is
a compact simple Lie group $G$,
with generators represented by matrices $\tau_a$
obeying the algebra $[\tau_a,\tau_b]$ $=$ $if_{abc}\tau_c$,
with a normalized trace $\tr\tau_a\tau_b= \d_{ab}$. The gauge real
superfield field $\f$ -- the prepotential -- is written in matrix form:
$\f=\tau_a\f^a$, as well as the Lagrange multiplier, ghost
and antighost chiral superfields $B$, $c_+$ and $c_-$, as well as
their conjugates $\Bb$, $\cb_+$ and $\cb_-$.
Matter is described by a set of chiral superfields $A^i$ belonging to some
-- may be reducible -- representation of the gauge group $G$,
the generators being represented
by Hermitean matrices $T_a{}^i{}_j$. The BRS transformations read
\eq\ba{ll}
s\f = \half \lc\f,c_++\cb_+\rc + M(\f)\lp c_+ -\cb_+\rp & \es
\phantom{s\f} = c_+-\cb_+ +\half\lc\f,c_++\cb_+\rc
  + \dfrac{1}{12} [\f,[\f, c_+-\cb_+]] + &\cdots\ ,\es
sc_+ = -\half\lac c_+,c_+\rac\ , \quad
&s\cb_+ = -\half\lac \cb_+,\cb_+\rac\ , \es
sA = -c_+^a T_a A \ ,\quad &s\Ab = \cb_+^a T_a \Ab \ ,\es
sc_- = B\ ,\quad& s\cb_- = \Bb\ ,\es
sB = 0\ ,\quad& s\Bb=0\ .
\ea\eqn{brs}
A concise notation has been used. For example, the first line means
\eq
s\f^a = \dfrac{i}{2} f_{abc} \f^b(c_++\cb_+)^c
  + M_{ab}(\f) (c_+ -\cb_+)^b\ ,
\eqn{brs-phi}
where the matrix $M_{ab}$ may be computed as a power series in $\f$
from the usual definition of the BRS transformation of the
prepotential $\f$ (written in matrix notation):
\eq
s e^\f = e^\f c_+ - \cb_+e^\f\ .
\eqn{brs-exp}
It is a remarkable fact that, except the term linear in $\f$, all
the contributions to the BRS transformation of $\f$ depend only
on the difference\footnote{This feature, i.e. the  structure
of \equ{brs-phi} is still true if one replaces
$\f$ by any Lie algebra valued function $\FF(\f)$ in \equ{brs-exp}, this
leading to a most general~\cite{ps-book} transformation law for $\f$.}
$(c_+-\cb_+)$.
In the supersymmetric Landau gauge, defined by
\eq
\dfud{\S}{B} =   \dfrac{1}{8}\Db^2D^2\f\ , \quad
\dfud{\S}{\Bb} =   \dfrac{1}{8}D^2\Db^2\f \ ,
\eqn{gauge-cond}
the complete invariant classical action $\S$ reads
\eq
\S = \S_{\rm gauge\ inv.} + \S_{\rm gauge\ fixing} + \S_{\rm ext.\ fields}\ ,
\eqn{action}
where
\[\ba{l}
\S_{\rm gauge\ inv.} = -\dfrac{1}{128g^2} \tr\ds F^\a F_\a
+ \dfrac{1}{16} \dv \Ab e^{\f^a T_a}A + \ds U(A)
  + \dsb \bar U(A)\ ,
\ea\]
\[\ba{l}
\S_{\rm gauge\ fixing} = s\lc \dfrac{1}{8} \tr\dv
      \lp c_- D^2\f + \cb_- \Db^2\f\rp \rc \es
\phantom{\S_{\rm gauge\ fixing}}
 = \dfrac{1}{8}\tr\dv \lp B D^2\f + \Bb \Db^2\f)
 - c_- D^2s\f -\cb_- \Db^2s\f \rp\ ,
\ea\]
\[
\S_{\rm ext.\ fields} = \dv\tr \f^*s\f + \lc \ds\lp A^{*i}sA_i
  +\tr c_+^* s c_+\rp    + {\rm c.c.} \rc\ .
\]
$F_\a$ denotes the supersymmetric
field strength $\Db^2(e^{-\f}D_\a e^\f)$,
and $U(A)$ the superpotential -- an invariant polynomial of degree 3 in
$A$. Moreover, $\f^*$, $A^*$ and $c_+^*$ are the external fields coupled to the
BRS transformations of $\f$, $A$ and $c_+$, respectively.

Denoting by $\G(\f,A,c_+,B,\f^*,A^*,c^*_+)$ the generating functional
of the 1PI Green functions, which coincide with the classical action
\equ{action} in the tree graph approximation, the BRS invariance of the
theory is expressed by the Slavnov-Taylor identity\footnote{We assume the
absence of gauge anomaly.}
\eq
\SS(\G) := \tr \dint d V {\dfrac{\d \G}{\d \f^*}} {\dfrac{\d \G}{\d \f}}
+ \lp \dint d S \left\{ {\dfrac{\d \G}{\d A^*_i}} {\dfrac{\d \G}{\d A_i}}
+ \tr {\dfrac{\d \G}{\d c_+^*}} {\dfrac{\d \G}{\d c_+}} + \tr B
{\dfrac{\d \G}{\d c_-}} \right\} + {\rm c.c.} \rp  = 0\ .
\eqn{slavnov}
The nilpotent
linearized operator associated to the Slavnov-Taylor operator at
$\G$ reads
\eq\ba{rl}
\SS_\G =& \tr \dint d V \lp
{\dfrac{\d \G}{\d \f^*}} {\dfrac{\d}{\d \f}} +{\dfrac{\d \G}{\d
\f}} {\dfrac{\d}{\d\f^*}} \rp    \es
&+\lp \dint d S \lp {\frac{\d \G}{\d
A^*}} {\dfrac{\d}{\d A}} + {\dfrac{\d \G}{\d A}} {\dfrac{\d}{\d A^*}}
+ \tr {\dfrac{\d \G}{\d c^*_+}} {\dfrac{\d }{\d c_+}} + \tr
{\dfrac{\d \G}{\d c_+}} {\dfrac{\d}{\d c^*_+}}
  + \tr B\dfud{}{c_-}\rp + {\rm c.c.}\rp  \ .
\ea\eqn{lin-slavnov}

\section{The Supersymmetric Antighost Equation}
In order to derive the classical form of the supersymmetric
antighost equation, let us differentiate the classical
action \equ{action} with respect to the ghost field $c_+$:
\eq\ba{l}
\dfud{\S}{c_+} = \dfrac{1}{16} \Db^2 [D^2c_-,\f] +
   \dfrac{1}{16}\Db^2 [\Db^2\cb_-,\f]
  - \dfrac{1}{2} \Db^2[\f^*-\frac{1}{8}(D^2c_- +\Db^2\cb_-),\f]
\es \phantom{\dfud{\S}{c_+}=}
    - \Db^2 \lp (\f^*-\frac{1}{8}(D^2c_- +\Db^2\cb_-))M(\f)\rp
    + [c_+^*,c_+] + A^*T_a A\tau_a \ .
\ea\eqn{deriv-c}
At this point one should observe that the right-hand side of \equ{deriv-c},
besides terms which are linear in the quantum fields,
also contains nonlinear terms due to the presence of the
formal power series $M(\f)$ entering the BRS transformation \equ{brs}
of the gauge superfield. These composite terms, being subject to
renormalization, spoil the usefulness of this equation. However, considering
the corresponding equation for $\cb_+$:
\eq\ba{l}
\dfud{\S}{\cb_+} = \dfrac{1}{16} D^2 [\Db^2\cb_-,\f] +
   \dfrac{1}{16}D^2 [D^2c_-,\f]
  - \dfrac{1}{2} D^2[\f^*-\frac{1}{8}(D^2c_- +\Db^2\cb_-),\f]
   \es \phantom{\dfud{\S}{\cb_+}=}
    + D^2 \lp (\f^*-\frac{1}{8}(D^2c_- +\Db^2\cb_-)) M(\f)\rp
    + [\cb_+^*,\cb_+] - \Ab^*T_a \Ab\tau_a \ ,
\ea\eqn{deriv-cbar}
adding together the superspace integrals of the equations
\equ{deriv-c}, \equ{deriv-cbar} and using\footnote{Use has been made
of the identity
\[
\ds\Db^2[D^2c_-,\f] = \ds[c_-, \Db^2D^2\f]\ ,
\]
and of its complex conjugate.}
 the Landau gauge conditions \equ{gauge-cond},
one obtains the antighost equation we are looking for:
\eq
\GG_-\S = \D_{\rm class}\ ,
\eqn{class-antig-eq}
with
\eq
\GG_-:= \ds \lp \dfud{}{c_+} -\lc c_-,\dfud{}{B} \rc\rp
  + \dsb \lp \dfud{}{\cb_+} -\lc \cb_-,\dfud{}{\Bb} \rc\rp
\eqn{g-minus}
and
\eq
\D_{\rm class} := -\dv [\f^*,\f]
  + \ds\lp [c_+^*,c_+] + (A^*T_a A)\tau_a \rp
  + \dsb\lp [\cb_+^*,\cb_+] - (\Ab T_a \Ab^*)\tau_a \rp\ .
\eqn{class-break}
We remark that the undesired nonlinear terms present in each of
the equations \equ{deriv-c} and \equ{deriv-cbar} have been cancelled.
We are thus left with the breaking \equ{class-break} which,
being now linear in the quantum fields, will not be renormalized,
i.e., it will remain a classical breaking.

Equation \equ{class-antig-eq} has now a form which allows one
to consider its validity to all orders of perturbation theory.
That it indeed holds as it stands at the quantum level:
\eq
\GG_-\G = \D_{\rm class}\ ,
\eqn{antig-eq}
 may be shown without any difficulty by repeating exactly the
argument given in~\cite{bps,p-sor-book} for the nonsupersymmetric case.

Let us finally remark that the sum of the superspace-integrated functional
derivatives with respect to $c_+$ and $\cb_+$ in \equ{g-minus} is in
fact the space-time integral of the functional derivative with respect
to the real part of the $\theta=0$ component of $c_+$. It coincides with
the functional operator appearing in the nonsupersymmetric
version of the antighost equation.
\subsection*{Rigid Invariance:}
Using the ``anticommutation relation''
\[
\GG_-\SS(\G) + \SS_\G\D_{\rm class} =
 -\ds\lp \lac c_-,\dfud{\G}{c_-}\rac + \lc B,\dfud{\G}{B}\rc\rp
 -\dsb\lp \lac \cb_-,\dfud{\G}{\cb_-}\rac +
                  \lc \Bb,\dfud{\G}{\Bb}\rc\rp\ ,
\]
one easily checks that the following identity holds:
\eq\ba{l}
W_{\rm rigid}\G := \dv\lp \lc \f,\dfud{\G}{\f}\rc
   + \lac \f^*,\dfud{\G}{\f^*}\rac\rp \es
+\ds\lp \lac c_+,\dfud{\G}{c_+}\rac + \lc c^*_+,\dfud{\G}{c^*_+}\rc
      + \lc B,\dfud{\G}{B}\rc + \lac c_-,\dfud{\G}{c_-}\rac
      +\lp\dfud{\G}{A} T_a A\rp\tau_a
             - \lp A^* T_a \dfud{\G}{A^*}\rp\tau_a \rp \es
+\dsb\lp \lac \cb_+,\dfud{\G}{\cb_+}\rac + \lc \cb^*_+,\dfud{\G}{\cb^*_+}\rc
      + \lc \Bb,\dfud{\G}{\Bb}\rc + \lac \cb_-,\dfud{\G}{\cb_-}\rac
      -\lp\Ab T_a\dfud{\G}{\Ab}\rp\tau_a
        - \lp \dfud{\G}{\Ab^*} T_a \Ab^* \rp\tau_a \rp \es
 = 0 \ .
\ea\eqn{wi-rig}
This is the Ward identity expressing the invariance of the theory
under the rigid transformations, corresponding to the
transformations of the gauge group,
but with constant (superspace independent) parameters.

\section{Nonrenormalization Theorems for the Chiral\hfill\break Anomalies}
This theorem is stated in general terms and proven
in~\cite{lps-beta-hpa} as Theorem A.1. Its main consequence
is Corollary A.2 of~\cite{lps-beta-hpa}, which may be paraphrased as follows.

Let the infinitesimal linear transformation\footnote{These
transformations commute with supersymmetry.
The nonrenormalization theorem in fact also holds for the
Fayet $R$ symmetry~\cite{lps-beta-hpa}.}
\eq
\d A^i = i e^i{}_j A^j\ ,\quad \d\Ab_i = -i\Ab_j e^j{}_i\ ,
\eqn{chiral-trf}
acting on the matter chiral superfields --
the numbers $e^i{}_j$ being the elements of an hermitian matrix
which commutes with the gauge group generators $T_a$ -- be a symmetry
of the theory up to a possible soft breaking due to the mass terms.
Such a symmetry is renormalizable~\cite{p-sor-book} and may
be expressed by a Ward identity
\eq
W\G := e^i{}_j \lc \ds A^j\dfud{}{A^i} - \dsb \Ab_i\dfud{}{\Ab_j}\rc\G \sim 0\
,
\eqn{rigid-wi}
where $\sim$ means equality up to soft breakings.

Then there exists a BRS invariant ``current'' $J_{\rm inv}$ -- a real
scalar superfield insertion -- such that
\eq
\Db^2 J_{\rm inv}\cdot\G = e^i{}_j A^j\dfud{\G}{A^i} + r \Db^2K^0\cdot\G\ .
\eqn{anom-conserv}
Without the last term, which will be explained below, this would be just
the Ward identity expressing the conservation of the current
represented by the vector component of the superfield $J_{\rm inv}$,
i.e. of the Noether current associated to the invari
ance under the transformation \equ{chiral-trf}.
the last term is an anomaly since it cannot be reabsorbed in a BRS
invariant way as a counterterm to $J_{\rm inv}$.
Indeed $K^0$ is a solution -- unique up to trivial terms -- of the
following supersymmetric version of the descent equations:
\eq\ba{l}
\ssg[K^0\cdot\G] = \Db_\ap[K^{1\ap}\cdot\G]\ ,\es
\ssg[K^{1\ap}\cdot\G] = (\Db^\ap D^\a+2D^\a\Db^\ap)[K^2_\a \cdot\G]\ ,\es
\ssg[K^2_\a\cdot\G] = D_\a[K^3\cdot\G]\ ,\es
\ssg[K^3\cdot\G] = 0\ ,
\ea\eqn{desc-eq}
whose solution is a quantum extension of the following superfield polynomials:
\eq\ba{l}
K^0 = \tr(\vf^\a\Db^2\vf_\a)\ ,\es
K^{1\ap} = -\tr(D^\a c_+\Db^\ap\vf_\a + \Db^\ap D^\a c_+\vf_\a)\ ,\es
K^2_\a = \tr(c_+D_\a c_+)\ ,\es
K^3 = \dfrac{1}{3}\tr c_+^3\ ,
\ea\eqn{sol-desc-eq}
where we have introduced the chiral superconnection
\[
\vf_\a= e^{-\f}D_\a e^\f\ .
\]
The main statement of the nonrenormalization theorem is that
the anomaly coefficient $r$ in \equ{anom-conserv} is exactly
given by its one-loop approximation.

The proof, which we shall not repeat, is based on the finiteness of
the insertion $K^3$ -- which is equivalent to the vanishing of its
anomalous dimension. The latter property was shown in~\cite{lps-beta-hpa}
as a consequence of the nonrenormalization theorem for the chiral
vertices~\cite{chiralvert}.
But, since the latter theorem requires exact supersymmetry and exact
superspace Feynman rules, and since supersymmetry is broken
in some way in most cases of interest, an alternative for the
finiteness of $K^3$, generalizing the one given in~\cite{bps} for
the nonsupersymmetric case, will be given now  using the
antighost equation \equ{antig-eq}. This proof is purely
algebraic~\cite{p-sor-book} and thus covers all the situations,
in paticular those whith a supersymmetry breaking described
in terms of superfields shifted by
$\theta$-dependent, $x$-independent quantities~\cite{ps-book,mop}.

The proof of the finiteness of the ghost monomial $\tr c_+^3$
follows exactly the one given in~\cite{bps,p-sor-book} for the
nonsupersymmetric theories. We couple it to an external chiral
superfield $\eta$, of dimension 3 and ghost number $-3$, i.e. we add
to the action the term
\eq
\dfrac{1}{3}\ds\eta\tr c_+^3 - \dfrac{1}{3}\dsb\bar\eta\tr \cb_+^3\ .
\eqn{eta}
The total action is still BRS invariant, the
Slavnov-Taylor identity, the gauge condition and the ghost equation
remain unchanged. The antighost equation stays as in \equ{antig-eq},
with the same classical breaking, but with the modified differential operator
\eq
\GG_-= \ds \lp \dfud{}{c_+} -\lc c_-,\dfud{}{B} \rc - \eta\dfud{}{c_+^*}\rp
  + \dsb \lp \dfud{}{\cb_+} -\lc \cb_-,\dfud{}{\Bb} \rc
    + \bar\eta\dfud{}{\cb_+^*}\rp\ .
\eqn{g-minus-eta}
Let us now look at the possible $\eta$-dependent invariant counterterms
$\D$ one may add to the action -- i.e. the counterterms corresponding
to a renormalization of $\tr c_+^3$. The only possible one is \equ{eta}
itself, which however is ruled out by the condition
\eq
\GG_-\D=0 \ .
\eqn{g-delta}
This concludes the algebraic proof of the nonrenormalization
theorem  of the anomaly of the type corresponding to
the descent equations \equ{desc-eq}.

The finiteness of the higher invariant ghost monomials
$\tr c_+^{2p+1}$, $p\ge2$, as well as of the ghost field itself,
goes along the same lines, as shown e.g. for $\tr c_+^5$ in
usual gauge theories~\cite{c-cinq} in the proof of the nonrenormalization of
 the gauge anomaly.

\vspace{8mm}

\noindent {\bf Acknowledgments}:  We are very indebted to the Conselho Nacional
de Pesquisa e Desevolvimento
(Brazil)  for its financial support, and we are grateful for interesting
discussions with
M.A. de Andrade, O.M. del Cima and M.W. de Oliveira.
One of the authors (O.P.) would like to thank the Institute of Physics of
the Catholic University of Petr\'opolis (Brazil) and its head Prof. R. Doria,
as well as the CBPF (Rio de Janeiro),
where part of this work has been done, for their hospitality.


\end{document}